# On the vacuum fluctuations, Pioneer Anomaly and Modified Newtonian Dynamics

Dragan Slavkov Hajdukovic[1]
PH Division CERN
CH-1211 Geneva 23;
dragan.hajdukovic@cern.ch
[1]On leave from Cetinje; Montenegro

Abstract
We argue that the so-called "Pioneer Anomaly" is related to the quantum vacuum fluctuations. Our approach is based on the hypothesis of the gravitational repulsion between matter and antimatter, what allows considering, the virtual particle-antiparticle pairs in the physical vacuum, as gravitational dipoles. Our simplified calculations indicate that the anomalous deceleration of the Pioneer spacecrafts could be a consequence of the vacuum polarization in the gravitational field of the Sun. At the large distances, the vacuum polarization by baryonic matter could mimic dark matter what opens possibility that dark matter do not exist, as advocated by the Modified Newtonian Dynamics (MOND).

According to Quantum Field Theory (QFT), the physical vacuum is a "kingdom" of virtual particle-antiparticle pairs. Some of them, like electron-positron pairs, may be considered as the virtual electric dipoles. Consequently, there is the phenomenon of the vacuum polarization by an external electric field.

In the present paper, we put forward the hypothesis that the quantum vacuum contains also the virtual gravitational dipoles. Consequently, the vacuum polarization by an external gravitational field should exist as well. In the framework of this hypothesis we suggest that the so called "Pioneer Anomaly", i.e. a small, constant sunward deceleration, of spacecrafts Pioneer 10 and Pioneer 11 (See Turyshev and Toth (2010) for a Review), could be a consequence of the vacuum polarization caused by the gravitational field of the Sun.

The simplest and the most economical possibility is to identify the virtual gravitational dipoles with the virtual particle-antiparticle pairs of the Standard Model of particle physics; hence, there is no need to introduce new, exotic particles, of unknown nature. Identification of the gravitational dipoles with particle-antiparticle pairs is possible only if antiparticles have a negative gravitational mass. Hence, instead of the current wisdom $\overline{m}_g = m_g$, we assume $\overline{m}_g = -m_g$, where a symbol with a bar denotes antiparticles, while $g$ refers to the gravitational mass. Let us point that we have introduced the negative gravitational mass (it should be better to say the negative gravitational charge) for antiparticles, while the inertial mass, relevant for non-gravitational phenomena, stays positive.

If the two gravitational masses $m$ and $-m$ are separated by a distance $\vec{d}$ pointing in the direction of the positive mass, the gravitational dipole moment is defined as

$$\vec{p}_g = m\vec{d} \tag{1}$$

In addition, it is appropriate to define the gravitational polarization density $\vec{P}_g$ as the gravitational dipole moment per unit volume.

In QFT, a virtual particle-antiparticle pair (i.e. a gravitational dipole) occupies the volume $\lambda_m^3$, where $\lambda_m$ is the corresponding Compton wavelength. Hence, the number density of the virtual gravitational dipoles has constant value

$$N_0 \approx \frac{1}{\lambda_m^3} \tag{2}$$

Let us point out that $\lambdabar_m$ can be interpreted as the mean distance between two dipoles which are the first neighbours. The size $d$ of a gravitational dipole (i.e. the mean distance between particle and antiparticle in a pair) should be significantly smaller than $\lambdabar_m$. In the absence of an accurate estimate for $d$, it is useful to observe that the size of a dipole and the reduced Compton wavelength ($\lambdabar_m \equiv \hbar/mc$) should have the same order of the magnitude ($d \sim \lambdabar_m$). Consequently,

$$p_g = md \sim \frac{\hbar}{c} \tag{3}$$

Hence, $\hbar/c$ is a universal order of magnitude, for all virtual dipoles.

In absence of an external gravitational field, the polarization density $P_g$ is zero. A sufficiently strong gravitational field (i.e. when the acceleration $a$ is greater than a critical value $a_{cr}$), can force all dipoles to be aligned along the field. If so, according to (2) and (3), the norm of $\vec{P}_g$ must have a maximum value

$$\left|\vec{P}_g\right| \equiv P_g = \frac{A}{\lambdabar_m^3}\frac{\hbar}{c}; \quad a > a_{cr} \tag{4}$$

where $A$ should be a dimensionless constant of the order of unity.

A simple estimate of the order of magnitude of $a_{cr}$ is

$$a_{cr} = \frac{Gm}{\lambdabar_m^2} \equiv \left(\frac{c}{h}\right)^2 Gm^3 \tag{5}$$

i.e. the gravitational acceleration produced by a particle at the distance of its own Compton wavelength.

If the quantum vacuum "contains" the virtual gravitational dipoles, a massive body with mass $M_b$ (a star, a black hole...), but also multi-body systems as galaxies should produce vacuum polarization, characterized with a gravitational polarization density $\vec{P}_g$.

As well known, in a dielectric medium the spatial variation of the electric polarization generates a charge density $\rho_b = -\nabla \cdot \vec{P}$, known as the bound charge density. In an analogous way, the gravitational polarization of the quantum vacuum should result in a gravitational mass density of the vacuum:

$$\rho_v = -\nabla \cdot \vec{P}_g \tag{6}$$

In the case of a star like the Sun, we may assume the spherical symmetry, and the equation (6) reduces to

$$\rho_v(r) = \frac{1}{r^2}\frac{d}{dr}[r^2 P_g(r)]; \quad P_g(r) \geq 0 \tag{7}$$

In principle there is a critical radius $R_{cr}$ (which will be determined later), so that for $r < R_{cr}$, all gravitational dipoles should be aligned along the gravitational field. In this region $P_g(r)$ must have a constant value determined by (4). Hence, equation (7) leads to

$$\rho_v(r) = \frac{2A}{\lambdabar_m^3}\frac{\hbar}{c}\frac{1}{r}; \quad r < R_{cr} \tag{8}$$

The corresponding mass enclosed in the sphere with radius $r$ is

$$M_v(r) = \frac{4\pi A}{\lambdabar_m^3}\frac{\hbar}{c}r^2; \quad r < R_{cr} \tag{9}$$

Now, the acceleration is given by the Newton law

$$a(r) = -\frac{G}{r^2}(M_b + M_v) = -\frac{GM_b}{r^2} - \frac{4\pi AG}{\lambdabar_m^3}\frac{\hbar}{c}; \quad r < R_{cr} \tag{10}$$

The equation (10) contains the anomalous sunward deceleration

$$a_P = -\frac{4A\pi G}{\lambda_m^3}\frac{\hbar}{c}; \quad r < R_{cr} \tag{11}$$

just as needed for the explanation of the Pioneer Anomaly. However it remains to specify $\lambda_m$, $A$ and $R_{cm}$. In particular it is necessary to show, that during the measurements, the distance of spacecrafts from the Sun was smaller than $R_{cr}$.

Recently, two independent approaches (Urban and Zhitnitsky 2009 and Hajdukovic 2010a, 2010b, 2010c) have supported the point of view that only QCD (quantum chromodynamics) vacuum is significant for gravitation. Roughly speaking the QCD vacuum is a gas of virtual pions (quark-antiquark pairs); consequently $\lambda_m$ in equation (11) should be identified with the Compton wavelength of pion ($\lambda_m = \lambda_\pi$). This identification, together with the choice $A = 2$, leads to

$$a_P = -8\pi G \frac{1}{\lambda_\pi^3}\frac{\hbar}{c} \tag{12}$$

what is in an intriguing numerical agreement with the observed anomalous acceleration of Pioneer spacecrafts. Using the mass of $\pi^\pm$, equation (12) gives

$$a_P = -8.4 \times 10^{-10} \, m/s^2$$

while the observed acceleration is

$$a_P = -(8.74 \pm 1.33) \times 10^{-10} \, m/s^2$$

It remains to determine $R_{cr}$ corresponding to $a_{cr}$. The condition of equality between $a_{cr}$, determined by equation (5), and the Newtonian acceleration $GM_b/R_{cr}^2$ leads to

$$R_{cr} = \lambda_\pi \sqrt{\frac{M_b}{m_\pi}} \tag{13}$$

For the Sun, the numerical result is $R_{cr} \approx 5000\, au$, what is much larger than the distances $r < 100\, au$ at which spacecrafts were observed. Hence, the measurements were done in the region of the constant gravitational polarization density.

According to the equations (12) and (13), the anomalous acceleration does not depend on the mass $M_b$. It has the same constant value for all bodies. However, the critical radius $R_{cr}$, i.e. the size of the space in which the phenomenon exists, increases as $\sqrt{M_b}$.

If the Pioneer anomaly has a gravitational origin, the additional acceleration should also affect the orbital motions of the Solar system's bodies. It is important to note that the ordinary Newtonian gravitational acceleration, caused by the Sun is much larger than the anomalous acceleration (12). Hence, the anomalous acceleration causes only a small perturbation of the Newtonian orbits. Two gravity models (without and with an anomalous acceleration) predict slightly different orbits, but it is still an open question (Page et al 2009, Iorio 2010) if the accuracy with which we know the orbits is sufficient to discriminate between these two cases.

Let us end with a few comments.

For $r < R_{cr}$ (i.e. $a > a_{cr}$) the polarization density is constant what results in an anomalous constant acceleration towards the central body.

For $r > R_{cr}$ (i.e. $a < a_{cr}$), the external gravitational field is not sufficiently strong to align all dipoles; hence the polarization density $P_g(r)$ should decrease with distance $r$. In the particular case when $P_g(r)$ decreases as $1/r$, the equation (7) leads to the conclusion that the gravitational mass density of the vacuum, $\rho_v(r)$, decreases as $1/r^2$ what is exactly the same as the mass distribution of the hypothetical dark matter. Hence, in the region $r > R_{cr}$ ($a < a_{cr}$), the polarization density could mimic dark matter what opens possibility that dark matter do not exist; the phenomenon might be consequence of the vacuum polarization by known baryonic matter. If it is so, the critical acceleration $a_{cr}$ should be the same as the acceleration $a_0$, considered as a new fundamental constant, in the framework of The Modified Newtonian Dynamics (MOND); for a Review of MOND see Milgrom 2008.

The numerical value of $a_0$ that fits the data is $a_0 \approx 1.2 \times 10^{-10} \, m/s^2$ and it is close to the Hubble scale $a_0 \approx cH_0$; what is a still unexplained numerical coincidence. On the other side, using mass of $\pi^{\pm}$ in the equation (5) gives $a_{cr} \approx 2 \times 10^{-10} \, m/s^2$, what is close to $a_0$. More than that, mass of a pion could be expressed by fundamental physical constants and cosmological parameters (Dirac 1937, 1938; Weinberg 1972; Hajdukovic 2010a, 2010b). Equation (5) together with relation (8) from Hajdukovic 2010a, leads to

$$a_{cr} = \frac{1}{4\pi^2} \frac{\Omega_{\Lambda 0}}{\sqrt{\Omega_0 - 1}} cH_0 \qquad (14)^*$$

where $\Omega_{\Lambda 0}$ and $\Omega_0$ are the usual dimensionless cosmological parameters corresponding to the present day values of the dark energy density (identified with the cosmological constant) and the total energy density of the Universe.

Apparently, the hypothesis of the gravitational repulsion between matter and antimatter might be a good starting point for understanding of the Pioneer Anomaly, but also it may help provide insight into the physical ground of the Modified Newtonian Dynamics.

*Important note

In the paper published in Astrophysics and Space Science, the factor $1/4\pi^2$ in equation (14) was omitted by mistake during the preparation of the manuscript. By the way, the right-hand side of equation (14) does not change with the expansion of the Universe (Hajdukovic 2010a. Hajdukovic 2010b)